\documentclass[letterpaper,12pt]{article}%
\usepackage{amsmath}
\usepackage{graphicx}
\usepackage{setspace}
\usepackage{amsfonts}
\usepackage{amssymb}%
\providecommand{\U}[1]{\protect\rule{.1in}{.1in}}
\begin{document}

\title{Pressure-Induced Superconductivity in Sc to 74 GPa}
\author{J.J. Hamlin and J.S. Schilling$\vspace{0.4cm}$\\\textit{Department of Physics, Washington University}\\\textit{\ CB 1105, One Brookings Dr, St. Louis, MO 63130}}
\maketitle

\begin{abstract}
Using a diamond anvil cell with nearly hydrostatic helium pressure medium we
have significantly extended the superconducting phase diagram $T_{c}(P)$ of
Sc, the lightest of all transition metals. We find that superconductivity is
induced in Sc under pressure, $T_{c}$ increasing monotonically to $8.2$ K at
74.2 GPa. The $T_{c}(P)$ dependences of the trivalent $d$-electron metals Sc,
Y, La, and Lu are compared and discussed within a simple $s\rightarrow d$
charge transfer framework.

\end{abstract}

\newpage

\section{Introduction}

Even though half a century has passed since the development of a microscopic
theory of superconductivity \cite{bcs}, it is still not possible to reliably
calculate values of the superconducting transition temperature $T_{c}$ for a
given material; in fact, one is not even able to reliably predict which
materials become superconducting and which do not. One strategy to make
progress in this situation is to establish systematics in $T_{c}$ as a
function of composition across alloy and compound series and then test whether
a particular theoretical approach is able to account for these systematics. A
related strategy is to look for systematics in the dependence of $T_{c}$ on
high pressure in a particular class of materials \cite{schilling2}. This
latter \textquotedblleft high-pressure\textquotedblright\ approach has the
advantage of being able to track changes in $T_{c}$ on a single sample, but
often has the disadvantage of being able to generate only relatively modest
changes in $T_{c}$. The use of the diamond-anvil cell alleviates this problem
since, by extending the pressure range to the multi-Megabar region, it is
capable of generating sizeable changes in the superconducting properties.

The $T_{c}(P)$ systematics in the simple-metal superconductors like Al, In, Sn
and Pb, where the conduction electrons possess $s,p$-character, are very
simple, namely, $T_{c}$ always decreases under pressure, i.e. the
superconductivity is weakened \cite{schilling2}. The reason for this is that
the pressure-induced changes in the lattice vibrations dominate over those in
the electronic system, leading to a decrease in $T_{c}$ as the lattice
stiffens under pressure. In such simple-metal systems it is an interesting
physics question to explore the manner in which $T_{c}$ approaches 0 K as the
pressure in increased; pioneering studies in this direction were carried out
in the 1970's by Gubser and Webb \cite{gubser1} on superconducting Al by
combining diamond-anvil cell, dilution refrigeration, and SQUID detection
technology \cite{gubser2}. Such high-pressure investigations at mK and sub-mK
temperatures, however, are extraordinarily difficult and require that the
materials studied be highly purified to contain only trace concentrations of
magnetic impurities.

Rather than use high pressures to \textit{destroy} superconductivity, as in
the simple metals, an alternative approach with perhaps greater promise is to
use high pressure to \textit{create} superconductivity, i.e. to focus
investigations on nonsuperconducting materials which require high pressures to
become superconducting. In such studies not only can the behavior of
superconductivity near 0 K be studied, but also the maximum attainable value
of $T_{c}$ for a given class of materials can be explored. Of the 52 known
superconducting elements, fully 23 only become superconducting if sufficient
pressure is applied \cite{schilling2}. Particularly interesting in this regard
are the alkali and noble metals, none of which superconduct at ambient
pressure. Since they are simple metals, pressure would be expected to weaken
the pairing interaction, so they should never become superconducting, no
matter how high the pressure. And yet both Cs \cite{wittig4,wittig5} and Li
\cite{shimizu1,struzhkin1,deemyad1} do superconduct at sufficiently high
pressures, $T_{c}$ for Li even reaching 15 - 20 K. Neaton and Ashcroft have
shown that the electronic structure of Li \cite{neaton1} and Na \cite{neaton2}
becomes increasingly non-free-electron-like as the volume available to the
conduction electrons outside the ion cores rapidly diminishes under very high
pressures. Cs, in fact, becomes a transition metal above $\sim$ 3 GPa as its
5$d$-band begins to fill through $s\rightarrow d$ transfer \cite{grover1}.
Similar considerations are expected to apply to the electronic structure of
many \textquotedblleft simple-metal\textquotedblright\ materials
\cite{neaton2}. In transition metal and rare earth systems it has been
appreciated for some time that the $d$-electron concentration $n_{d}$
generally increases under pressure and is mainly responsible for the
systematic progression of crystal structures under pressure exhibited by both
systems \cite{duthie1,pettifor1}.

Superconductivity is most likely to occur in those materials containing one or
more nonmagnetic transition metal (or $d$-electron) elements, notable
exceptions being the trivalent metals Lu, Y, and Sc. Why are these three
elements not superconducting at ambient pressure, whereas isoelectronic La is?
The answer may lie in the fact that they simply don't have a sufficient number
of $d$ electrons to support superconductivity; La, on the other hand, has more
$d$ electrons due to its significantly larger ion core \cite{duthie1}. The
assertion that Lu, Y, and Sc have an insufficient $d$-electron count for
superconductivity is supported by the fact that all 3$d$-, 4$d$-,
5$d$-transition metals in columns IV and V do superconduct at ambient
pressure, and those in column V with their greater $d$-electron count have
values of $T_{c}$ roughly $20\times$ higher. Increasing the $d$-electron
concentration in Lu, Y, and Sc by applying high pressure would be expected,
therefore, to promote superconductivity. Indeed, Wittig\textit{\ et al.} were
the first to show this to be true for Lu \cite{wittig6, wittig1}, Y
\cite{wittig4}, and Sc \cite{wittig2}. Whereas in La $T_{c}(P)$ passes through
a maximum near 13 K \cite{wittig3,tissen1}, that for Y continues to increase
to the highest pressure applied ($\simeq20$ K at 1.2 Mbar). $T_{c}$ for Lu
\cite{wittig1} and Sc \cite{wittig2} also increases under pressure, but only
reaches values of 2.5 K\emph{\ }and 0.35 K at 22 GPa and 21.5 GPa, respectively.

In this paper we extend the earlier studies \cite{wittig2} on elemental Sc to
much higher pressures. $T_{c}$ increases monotonically with pressure, reaching
8.2 K at 74.2 GPa. To help illuminate the nature of the superconductivity for
all four trivalent metals Sc, Y, Lu, and La, we search for systematics in the
dependence of $T_{c}$ on the free volume fraction available to the conduction electrons.

\section{Experiment}

The diamond anvil cell used contains two opposing 1/6-carat, type Ia diamond
anvils with 0.4 mm diameter culets. A miniature Sc sample ($\sim$ 70 $\mu$m
diameter $\times$ 35 $\mu$m thick) is cut from a high-purity ingot (99.98\%
metals basis) obtained from the Materials Preparation Center of the Ames
Laboratory \cite{ames1} and placed in a 180 $\mu$m diameter hole electro-spark
drilled through the center of a gold-sputtered NiMo gasket 3 mm in diameter by
250 $\mu$m thick and preindented to 45 $\mu$m thickness. Tiny ruby spheres
\cite{chervin_2001_1} are placed next to the Sc sample to allow the
determination of the pressure \textit{in situ} at 20 K with resolution $\pm$
0.2 GPa. We use the revised ruby pressure scale of Chijioke \textit{et al.}
\cite{silvera_2005_1}. The R1 ruby fluorescence line remains sharp up to the
highest pressures confirming the near hydrostaticity of the pressure
environment in the present experiment.

At the beginning of the experiment, the Sc sample and ruby spheres are placed
in the gasket hole. The pressure cell is then placed in a continuous flow
cryostat (Oxford Instruments) and submerged in liquid helium. To insure that
no bubbles of gaseous He are trapped inside the gasket, the helium is cooled
below the lambda point before sealing the high pressure volume by pressing the
diamonds into the gasket. At the highest pressures the Sc sample remained
completely surrounded by the nearly hydrostatic dense helium pressure medium.
To reduce the possibility of He diffusion into the diamond anvils, the
temperature was kept below 180 K during the entire experiment. Following the
initial compression at 1.6 K, the pressure was only changed between 100 K and
180 K.

The superconducting transition is detected inductively using a balanced
primary/secondary coil system connected to a Stanford Research SR830 digital
lock-in amplifier via a SR554 transformer preamplifier; the excitation field
for the ac susceptibility studies is 3 Oe r.m.s. at 1023 Hz. To facilitate the
recognition of the superconducting transition, a temperature-dependent
background signal $\chi_{b}^{\prime}\left(  T\right)  $ is subtracted from the
measured susceptibility data; $\chi_{b}^{\prime}\left(  T\right)  $ is
obtained by measuring at pressures too low to induce superconductivity. A
relatively low noise level is achieved by using the transformer preamplifier
to ensure good impedance matching, varying the temperature very slowly
(100~mK/min) at low temperatures, using a long time constant (30 s) on the
lock-in amplifier, and averaging over 2-3 measurements. Further experimental
details of the high pressure and ac susceptibility techniques are published
elsewhere \cite{hamlin1,schilling_1984_1,deemyad_2001_1}.

\section{Results and Discussion}

In Fig.~1 we show the results of the present ac susceptibility measurements
for nearly hydrostatic pressures from 54.3 to 74.2 GPa. The real part of the
ac susceptibility $\chi^{\prime}(T)$ decreases abruptly by 3-4 nV upon cooling
through the superconducting transition. $T_{c}$ is seen to increase
monotonically with pressure. Signal fluctuations arising from the $^{4}$He
boiling point and superfluid transition prevented the acquisition of reliable
data below 4 K. The shift in $T_{c}\simeq$ 8.2 K under an applied dc magnetic
field up to 500 Oe was less than the experimental resolution, implying that
$dT_{c}/dH\lesssim0.3$ mK/Oe. Since values of $dT_{c}/dH$ at low fields for
type I superconductors are typically a few mK/Oe, the superconductivity in Sc
is likely type II, as in La and Y. For an Y sample with $T_{c}\approx$ 9.7 K
at 46.6 GPa \cite{hamlin1}, $T_{c}$ was found to decrease under magnetic
fields to 500 Oe at the rate $dT_{c}/dH\approx-0.5$ mK/Oe.

In Fig.~2 the dependence of $T_{c}$ on pressure for Sc is shown from the
present experiment to 74.2 GPa and compared with the previous quasihydrostatic
pressure results of Wittig \textit{et al}. \cite{wittig2} to\ 21.5 GPa. It is
worth noting that, in contrast to the results for Y, the dependence of $T_{c}$
on pressure for Sc exhibits an upward (positive) curvature, in spite of the
fact that its compressibility \textit{decreases} with increasing pressure
\cite{grosshans5}. The accelerating increase in $T_{c}$ with pressure in Sc
gives hope that much higher values of $T_{c}$ can be reached in future
experiments in the multi-Megabar pressure range.

We now compare the change in $T_{c}$ under pressure from all known
high-pressure experiments on Sc, Y, La, and Lu. Instead of simply plotting
$T_{c}$ versus pressure, we plot in Fig.~3 $T_{c}$ versus the ratio
$r_{a}/r_{c}$ of the Wigner-Seitz radius $r_{a}$ to the ion core radius
$r_{c}$ \cite{springer1}. This ratio is directly related to the free volume
available to the conduction electrons outside the ion cores; the relative
decrease in this free volume under pressure is particularly rapid as the ion
cores draw close together and begin to overlap. Johansson and Rosengren
\cite{johansson} were the first to recognize that the ratio $r_{a}/r_{c}$
appears to play an important role in characterizing the pressure dependence of
$T_{c}$ in Y, La, Lu and La-Y, and La-Lu alloys as well as in the equilibrium
crystal structure sequence across the rare-earth series. Duthie and Pettifor
\cite{duthie1} subsequently demonstrated for La and Lu that the correlations
in the structure sequence are a consequence of the fact that the $d$-band
occupancy $n_{d}$ increases under pressure due to $s\rightarrow d$ transfer as
the equilibrium atomic volume decreases.

Although differing in detail, the $T_{c}$ versus $r_{a}/r_{c}$ data in Fig.~3
for Y, La, Sc, and Lu have important features in common, namely, that as
$r_{a}/r_{c}$ decreases under pressure, $T_{c}$ initially rises rapidly,
reaching $\sim$ 3-4 K for values of $r_{a}/r_{c}$ between 1.9 and 2.1. The
fact that superconductivity in Sc initiates at the relatively large ratio
$r_{a}/r_{c}\approx2.24$ fueled our interest in this metal since it suggested
to us that sufficient pressure might yield relatively high values of $T_{c}$.
With the exception of Sc, the similarities in the pressure dependences of
$T_{c}$ in Fig.~3 are matched by the similarities in the pressure-induced
changes in crystal structure \cite{vohra5,grosshans5} which fit in quite well
with the well-known hcp $\rightarrow$ Sm-type $\rightarrow$ dhcp $\rightarrow$
fcc structure sequence characteristic for the rare-earth metals. This is not
surprising since, with the exception of Eu and Yb, all rare earths are also
trivalent $d$-electron metals. Sc falls somewhat out of line since it
transforms at $\sim$ 23 GPa from the hcp to an incommensurate host-guest
structure \cite{fujihisa1,mcmahon1} instead of to the canonical Sm-type
structure. In fact, recent X-ray diffraction experiments on Sc to 297 GPa
reveal four successive structure changes, the final being to a new helical
chain structure above 240 GPa \cite{akahama1}. It has been suggested that the
differences between Sc and the other trivalent $d$-electron metals may arise
at least in part from the changes in electronic structure associated with the
complete absence of $d$-electrons in Sc's ionic core, thus allowing its 3$d$
valence electrons to penetrate further into the core region (no orthogonality
condition) and thus to assume a higher degree of localization
\cite{holzapfel10,olijnyk1}. The slow monotonic increase in the $E_{2g}$
vibration mode and the $C_{44}$ elastic shear modulus of Sc under pressure are
also anomalous \cite{olijnyk1}.

In Fig.~3 it is seen that the dependence of $T_{c}$ on $r_{a}/r_{c}$\ for Sc
matches rather well that for La but lies above those for Y and Lu. That the
$T_{c}$ versus $r_{a}/r_{c}$ dependences for these four trivalent $d$-metals
don't map on top of each other is not surprising. A more relevant parameter
for superconductivity than the ratio $r_{a}/r_{c}$ might be the number of
$d$-electrons per atom in the conduction band $n_{d}$. For La and Lu under
ambient conditions, for example, Duthie and Pettifor \cite{duthie1} estimate
that $n_{d}\simeq2.5$ and 1.9, respectively. Were the $T_{c}(P)$ versus
$n_{d}(P)$ dependences for these four elemental metals to fall closely
together, this would suggest that the simple $d$-electron count has a
particularly close tie to the superconductivity. It is, of course, clear that
a detailed understanding of $T_{c}(P)$ must necessarily take into account
pressure-induced changes in crystal structure. However, if we have learned
anything in the field of superconductivity, it is that real progress often
entails searching for and identifying overriding systematics. The available
data from experiment and theory are not yet sufficiently complete that a
possible correlation between $T_{c}$ and $n_{d}$ can be properly identified.
Still needed for this purpose are: \ (1) further $T_{c}(P)$ data on Y, Lu, and
Sc to much higher pressures, and (2) an accurate estimate of $n_{d}(P)$ for
all four trivalent $d$-elements from a unified electronic structure
calculation.\vspace{0.4cm}

\noindent Acknowledgments. \ The authors are grateful to R.W. McCallum and
K.W. Dennis of the Materials preparation Center, Ames Lab, for providing the
high purity Sc sample. Thanks are due V. Tissen for providing the NiMo gasket
material used in this experiment. The authors also gratefully acknowledge
research support by the National Science Foundation through grant DMR-0404505.

\begin{center}
{\LARGE Figure Captions}

\bigskip

\bigskip
\end{center}

\noindent\textbf{Fig.~1.} \ Real part of the ac susceptibility signal in
nanovolts versus temperature for Sc at different pressures ranging from 54.3
to 74.2 GPa. Curves are shifted vertically for clarity. The superconducting
transition temperature $T_{c}$, \ which is defined by the transition midpoint,
is seen to increase monotonically with pressure.\bigskip

\noindent\textbf{Fig.~2.} \ Superconducting transition temperature $T_{c}$
versus pressure to 74.2 GPa. Numbers give order of measurement. Dashed line is
guide to the eye and links present data ($\bullet$) to previous results of
Wittig \textit{et al.} to 21.5 GPa \cite{wittig2} (short solid line). \bigskip

\noindent\textbf{Fig.~3.} \ Superconducting transition temperature $T_{c}$
plotted versus ratio $r_{a}/r_{c}$ of Wigner-Seitz to ion core radius for
present data on Sc ($\bullet$) from Fig.~2, Y (solid line) from
Ref.~\cite{hamlin1}, Lu (solid line) from Refs.~\cite{wittig1, wittig3}, and
La (dotted line from Ref.~\cite{wittig3}, dot-dashed line from
Ref.~\cite{tissen1}. Vertical arrows mark values of $r_{a}/r_{c}$ for the
respective metal at ambient pressure \cite{springer1}. See Ref.
\cite{springer1} for full details regarding calculation of pressure dependence
of ratio $r_{a}/r_{c}$.

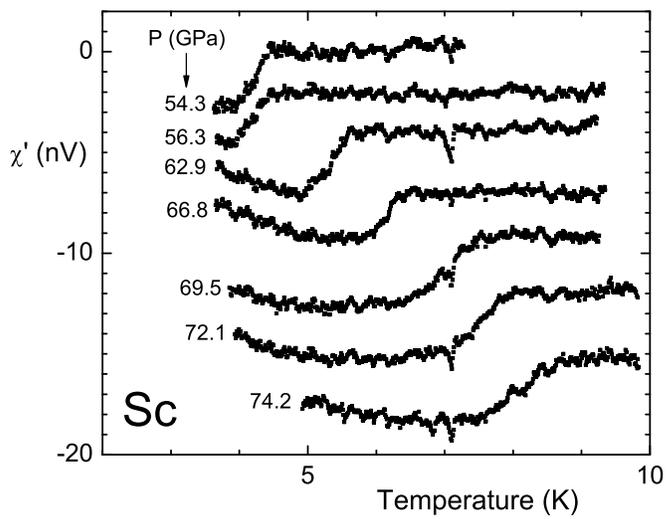
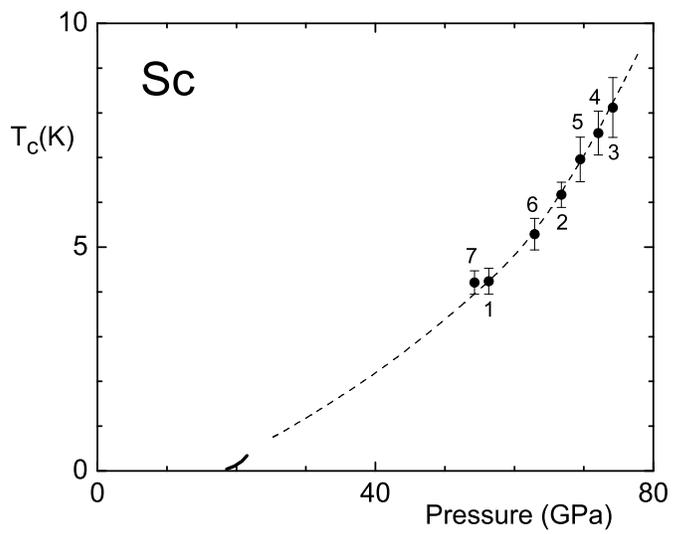
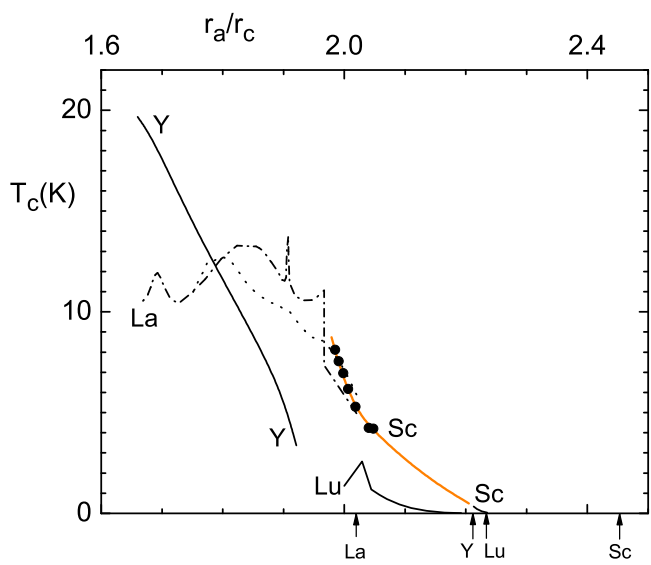
\end{document}